\documentclass[letterpaper]{article} 
\usepackage{aaai2027}  
\usepackage[hyphens]{url}  
\usepackage{graphicx} 
\usepackage{natbib}  
\usepackage{caption} 
\usepackage{tabularx}
\usepackage{amsmath,amssymb,amsfonts}
\usepackage{textcomp}
\usepackage{multirow}
\usepackage{booktabs}
\usepackage{textcase}
\usepackage{enumitem}
\usepackage{soul}
\usepackage{diagbox}
\usepackage{algpseudocode}
\usepackage{xspace}
\usepackage[most]{tcolorbox}
\usepackage[ruled,vlined,linesnumbered]{algorithm2e}
\usepackage{color}
\usepackage{xcolor}
\usepackage{listings}
\usepackage{tikz}
\usepackage{pgfplots}
\pgfplotsset{compat=1.18}
\usepgfplotslibrary{groupplots}
\usepackage{tikz-qtree}
\usepackage{varwidth}
\usetikzlibrary{calc,plotmarks}
\usetikzlibrary{patterns}
\usepackage{url}
\usepackage{subcaption}
\usepackage{threeparttable}
\usepackage{colortbl}
\usepackage{bbding}
\usepackage{placeins}

\newcommand{\affmark}[1]{\raisebox{0.7ex}{\scriptsize #1}}

\title{VulnGym: Benchmarking Coding Agents for Repository-Level Vulnerability Detection}
\author{
    Kexing Ji\affmark{1},
    Jiachen Liu\affmark{1},
    Enze Hu\affmark{1},
    Cuiyun Gao\affmark{2},
    Keke Lian\affmark{1},\\
    Yongheng Liu\affmark{3},
    Lei Zhang\affmark{3},
    Tian Dong\affmark{4},
    Hao Chen\affmark{4},
    Wang Bin\affmark{5}
}
\affiliations{
    \affmark{1}Tencent Wukong\\
    \affmark{2}The Chinese University of Hong Kong\\
    \affmark{3}Fudan University\\
    \affmark{4}The University of Hong Kong\\
    \affmark{5}Peking University\\
    kexing1208@gmail.com, jiachenliu1024@gmail.com, slamhu@tencent.com,\\
    cuiyungao@outlook.com, kekelian@tencent.com, yhliu24@m.fudan.edu.cn,\\
    zxl@fudan.edu.cn, nsectian@gmail.com, chenho@hku.hk,\\
    2201111747@stu.pku.edu.cn
}

\begin{document}

\maketitle

\begin{abstract}
Recent advances in LLM-based vulnerability detection have shown promising results, while coding agents further extend this capability from isolated code snippets to complete repositories.
This shift requires agents to autonomously explore repositories and locate vulnerability-relevant code, instead of performing detection on preselected functions.
However, existing benchmarks primarily focus on vulnerability classification over preselected code snippets, limiting their ability to evaluate coding agents in repository-level vulnerability detection.
Moreover, without fine-grained vulnerability trace annotations, the capability limitations underlying the detection process remain difficult to explore.
To address these limitations, we present \textbf{VulnGym}, a real-world repository-level benchmark for evaluating vulnerability detection by coding agents.
VulnGym aligns reviewed GitHub advisories with their corresponding vulnerable version repositories.
It contains 184 advisories and 408 vulnerability entries across 23 repositories, with each entry annotated with line-level entry points, critical operations, and vulnerability traces.
Using this fine-grained ground truth, VulnGym defines an end-to-end detection task and three oracle-based subtasks to jointly evaluate vulnerability detection and diagnose limitations in code localization and evidence construction.
Our evaluation indicates that current coding agents remain limited in both end-to-end repository-level vulnerability detection and the construction of accurate supporting traces.

\end{abstract}

\begin{sloppypar}
\section{Introduction} \label{sec:intro}
Large language models (LLMs) have shown promising performance in function-level vulnerability detection~\cite{zhou2025large, yildiz2025benchmarking, zhou2024large}.
Building on these advances, coding agents further combine iterative reasoning and planning with tool interaction~\cite{hong2024metagpt,yang2024sweagent,shinn2023reflexion}, enabling them to search files, trace code dependencies, and perform repository-level vulnerability detection~\cite{wang2024reposvul,yildiz2025benchmarking}.
This paradigm better aligns vulnerability detection with security practice through complete-repository analysis rather than preselected code snippets~\cite{li2025iris,wen2026function}.

However, this shift has exposed a substantial gap between existing vulnerability benchmarks and the requirements for evaluating code agents~\cite{islam2024mapcoder,zhang2024codeagent,qian2024chatdev,cursor2026,anthropic2025claudecode,yang2024sweagent,openhands2024} at the repository level.
These benchmarks generally focus on preselected functions, without extending the evaluation scope from located code to the broader repository~\cite{lian2026ase, yildiz2025benchmarking}.
In contrast, real-world vulnerabilities often span multiple functions, files, and modules.
For example, a stored XSS vulnerability~\cite{github2025cve64495} in Open WebUI~\cite{baek2026openwebui} propagates user-controlled prompt content through multiple components before reaching an unsafe \texttt{innerHTML} assignment.
Repository-level vulnerability detection requires coding agents to navigate the codebase and trace paths from entry points to unsafe operations to identify a vulnerability~\cite{wang2025cybergym}.

Recent work has broadened the evaluation scope from isolated functions to repository-level detection scenarios.
VulEval~\cite{wen2026function} augments target functions with repository-level dependency context and evaluates vulnerability classification under broader code contexts.
Similarly, JITVul~\cite{yildiz2025benchmarking} evaluates LLMs through binary classification of preselected functions extracted from paired vulnerable and patched repository versions.
Despite offering valuable insights into LLM-based vulnerability detection, these works share several key limitations for comprehensive benchmarking.
First, their evaluation remains centered on predefined target functions, so agents are not required to explore the broader repository and locate vulnerability-relevant code autonomously.
Second, their classification-oriented ground truth lacks fine-grained vulnerability-relevant traces, making in-depth analysis of detection behavior difficult.
Consequently, an agent may predict the correct label on given codes yet locate the wrong defect or rely on irrelevant evidence, while the evaluation cannot reveal where in the detection process the failure occurs.
Together, these limitations raise a critical question: \textbf{\textit{how can we evaluate and diagnose where the vulnerability detection process of coding agents fails in real-world repositories?}}

To answer this question, we introduce VulnGym, a benchmark for evidence-grounded vulnerability detection by coding agents in real-world repositories.
VulnGym contains 184 reviewed advisories and 408 vulnerability entries across 23 open-source repositories, with each entry aligned to its vulnerable repository version.
Rather than providing only vulnerability labels or predefined targets, VulnGym annotates each entry with line-level entry points, critical operations, and vulnerability traces.
These annotations enable both evaluating repository-level vulnerability detection and diagnosing where in the detection process an agent fails.
Based on this ground truth, VulnGym defines an end-to-end detection task and three oracle-based subtasks.
The end-to-end task jointly evaluates vulnerability findings and their supporting evidence, while the oracle-based subtasks diagnose limitations in locating relevant code and constructing traces.


Our main contributions are as follows:
\begin{itemize}[leftmargin=*,itemsep=1pt,topsep=2pt]
    \item We introduce VulnGym, a real-world repository-level benchmark with fine-grained code evidence annotations.
    It contains 184 reviewed advisories and 408 vulnerability entries across 23 open-source repositories.

    \item We design an evaluation framework comprising an end-to-end detection task and three oracle-based subtasks that jointly assess vulnerability detection and diagnose capability limitations in code localization and evidence construction.

    \item We conduct a systematic empirical study of current coding agents across model scales and LLM--agent combinations.
    Our findings reveal capability limitations underlying vulnerability detection.
\end{itemize}

\section{Related Work} \label{sec:backround}

Early vulnerability benchmarks formulate detection as function-level binary classification~\cite{lin2019deep}. BigVul~\cite{fan2020ac} and Devign~\cite{zhou2019devign} provide large labeled function collections but suffer from label noise and duplication~\cite{croft2021empirical,tamberg2025harnessing,liu2024vuldetectbench}. PrimeVul~\cite{ding2024vulnerability} improves labeling accuracy through automated strategies and removes exact duplicates using MD5 hashing, while SecVulEval~\cite{ahmed2026secvuleval} refines annotations to the statement level and extracts multiple forms of code context. These benchmarks still lack inter-procedural context and complete execution chains from entry points to critical operations.

VulEval~\cite{wen2026function} and CWE-Bench-Java~\cite{li2025iris} extend evaluation to repositories. VulEval localizes vulnerable files and lines from patches and provides cross-function call dependencies; CWE-Bench-Java contains 120 manually verified vulnerabilities in compilable Java projects, each linked to its fix commit. Both primarily evaluate classification accuracy, rather than autonomous repository exploration, and neither provides vulnerability execution chains.

JITVul~\cite{yildiz2025benchmarking} is the first benchmark of ReAct~\cite{yao2022react} agents for repository-level vulnerability detection. It frames detection as pairwise classification of vulnerability-introducing and fixing commits, lets agents retrieve cross-procedural context through tool calls, and proposes pAcc to avoid inflated F1 from over-predicting the vulnerable class. It nevertheless targets preselected functions and lacks chain-level annotations for assessing reasoning completeness.

\providecommand{\cmark}{\textcolor{black}{\Checkmark}}
\providecommand{\xmark}{\textcolor{black}{\XSolidBrush}}
\begin{table}[t]
  \centering
  {\small
  \setlength{\tabcolsep}{3pt}
  \begin{tabular}{l l ccc c}
    \toprule
    \textbf{Benchmark} & \textbf{Level}
      & \textbf{Location} & \textbf{Cause} & \textbf{Chain} & \textbf{Val.} \\
    \midrule
    PrimeVul        & Function        & \xmark & \xmark & \xmark & \xmark \\
    SecVulEval      & Function        & \cmark & \xmark & \xmark & \xmark \\
    VulEval        & Repository         & \cmark & \xmark & \xmark & \xmark \\
    CWE-Bench-Java  & Repository         & \xmark & \xmark & \xmark & \cmark \\
    JITVul          & Repository        & \xmark & \xmark & \xmark & \xmark \\
    \midrule
    \textbf{VulnGym} & Repository        & \cmark & \cmark & \cmark & \cmark \\
    \bottomrule
  \end{tabular}
  }
  \caption{%
    Comparison of vulnerability benchmarks.
    Level denotes the evaluation scope;
    Location denotes vulnerable file and line localization;
    Cause denotes a natural-language defect cause;
    Chain denotes a line-level execution trace across files;
    Val. indicates whether annotations are human-verified.
  }
  \label{tab:benchmark_comparison}
\end{table}

Table~\ref{tab:benchmark_comparison} summarizes existing benchmarks across five dimensions. No benchmark covers all five, particularly Chain and Cause, which are needed to evaluate autonomous repository exploration and complete evidence-chain construction. VulnGym fills these gaps with repository-level line annotations, natural-language defect causes, and human-verified cross-module Traces.

\section{VulnGym} \label{sec:VulnGym}


\begin{figure*}[t]
  \centering
  \includegraphics[width=0.95\textwidth]{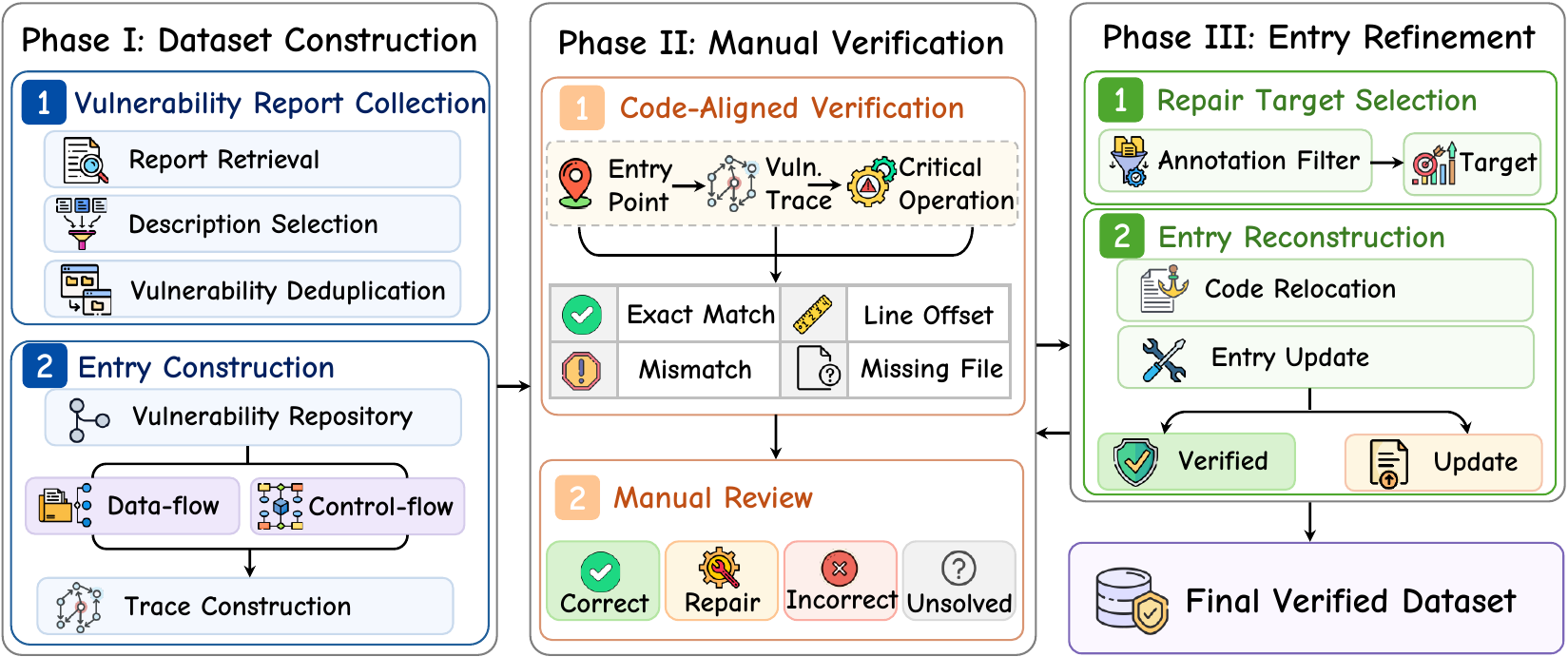}
  \caption{Overview of the VulnGym benchmark construction pipeline.}
  \label{fig:pipeline}
\end{figure*}

\subsection{Benchmark Construction}

As illustrated in Figure~\ref{fig:pipeline}, VulnGym is constructed in three phases: dataset generation, manual verification, and entry refinement. Phase I aligns advisories with their vulnerable repository snapshots and generates structured vulnerability records. Phase II verifies these records through source-code alignment and human review. Records that do not pass verification enter Phase III and return to Phase II after refinement, forming an iterative verification loop.

\paragraph{Phase I: Dataset Construction.}
We begin with 3,632 reviewed high risk severity GitHub Advisories\cite{github_advisory_database} published between January 2025 and April 2026, covering 1,579 open-source repositories.
To reduce potential data leakage when benchmarking coding agents\cite{deng2023benchmark,zhou2025lessleak}, we select advisories published after the training cutoff of LLM, from November 2025 to April 2026, and retain only repositories with more than 10,000 GitHub stars.
For each remaining advisory, we identify the corresponding vulnerable repository and establish its vulnerable snapshot using the detailed description and vulnerability databases.
Advisories whose repositories cannot be located or whose vulnerable snapshots cannot be reliably established are excluded, resulting in 184 advisories.

For each vulnerable repository snapshot, we construct reachable-entry records by identifying externally reachable Entry Points (EPs), security Critical Operations (COs), and the Traces connecting them through data-flow and control-flow analysis.
The EP is an externally reachable code location that can trigger the vulnerability, such as an API handler or event callback.
The CO is a security-critical location where the vulnerability takes effect, e.g., an unsafe operation or a missing security check.
The Trace is an ordered execution path connecting an EP to a CO, with each step corresponding to a specific file location and code snippet.
When multiple independent EP--CO paths exist, each is stored as a separate reachable-entry record.

\paragraph{Phase II: Manual Verification.}
This phase first compares the code locations associated with each EP, CO, and Trace step line by line against the target source code and classifies the result as an exact match, line offset, content mismatch, or missing file.
Two reviewers then independently verify the correctness of each entry based on the vulnerability description, including the external reachability of the EP, security relevance of the CO, and ordering of the Trace. 
Each entry is labeled as \texttt{Correct}, \texttt{Repair}, \texttt{Incorrect}, or \texttt{Unsolved}.
\texttt{Correct} denotes a fully valid entry.
\texttt{Repair} indicates an otherwise valid entry with localized errors in line numbers or the Trace, while \texttt{Incorrect} indicates that the entry misidentifies the vulnerability entirely.
\texttt{Unsolved} marks cases that reviewers cannot determine confidently and that require third-party adjudication.
When the two reviewers disagree on a label or an entry is labeled as \texttt{Unsolved}, a third senior reviewer adjudicates the case and makes the final decision.
We further conduct inter-rater agreement measurement using Cohen's Kappa~\cite{ma2024api}, with a score of $\kappa = 0.87$, indicating high agreement.
The review labels, together with annotations of erroneous line numbers and manual corrections for entries requiring changes, are then passed to Phase III for entry refinement.

\paragraph{Phase III: Entry Refinement.}
This phase selects repair targets based on the annotation labels and line-level correction feedback produced in Phase II.
For each selected entry, the refinement process first identifies the corresponding source files and code lines in the target repository.
It then updates the relevant entry fields based on the relocated code.
Entries requiring no revision after source-code validation are marked as \texttt{Verified}, whereas entries with source-grounded corrections are marked as \texttt{Update} and returned to Phase II for another round of verification.
Entries whose corresponding source files or line numbers cannot be reliably identified are excluded from the final verified dataset.

\begin{figure}[t]
  \centering
  \includegraphics[width=\linewidth]{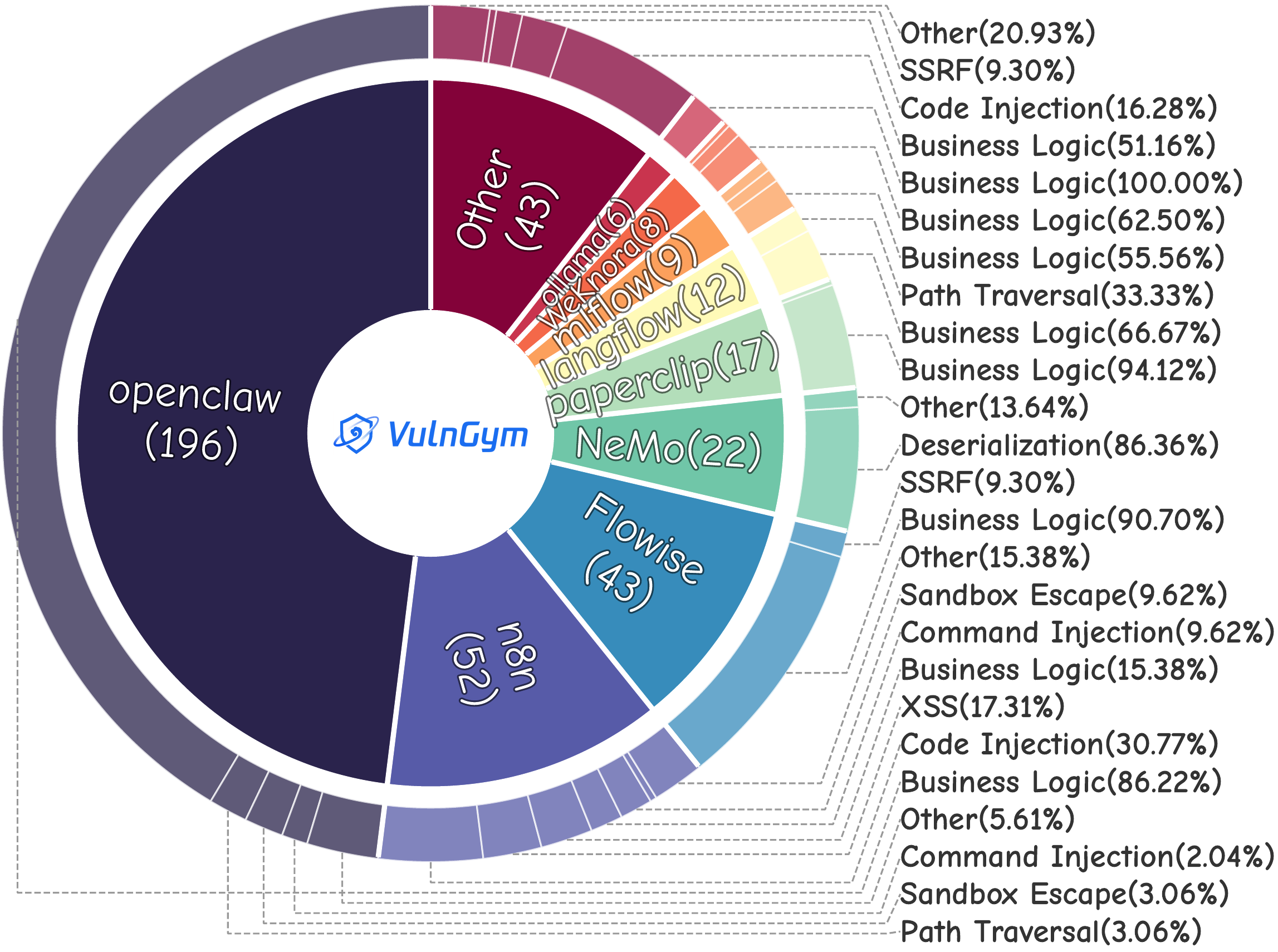}
  \caption{Repository-level composition of VulnGym.
    The inner ring shows the proportion of reachable-entry records contributed by
    each repository; the outer ring shows the distribution of vulnerability categories.}
  \label{fig:dataset_info}
\end{figure}

Following the pipeline above, we construct VulnGym with 408 reachable-entry records derived from 184 advisories across 23 open-source repositories.
Figure~\ref{fig:dataset_info} shows the number of entries from each repository and the corresponding distribution of vulnerability categories.

\subsection{Evaluation Framework}

\begin{figure}[t]
  \centering
  \includegraphics[width=\linewidth]{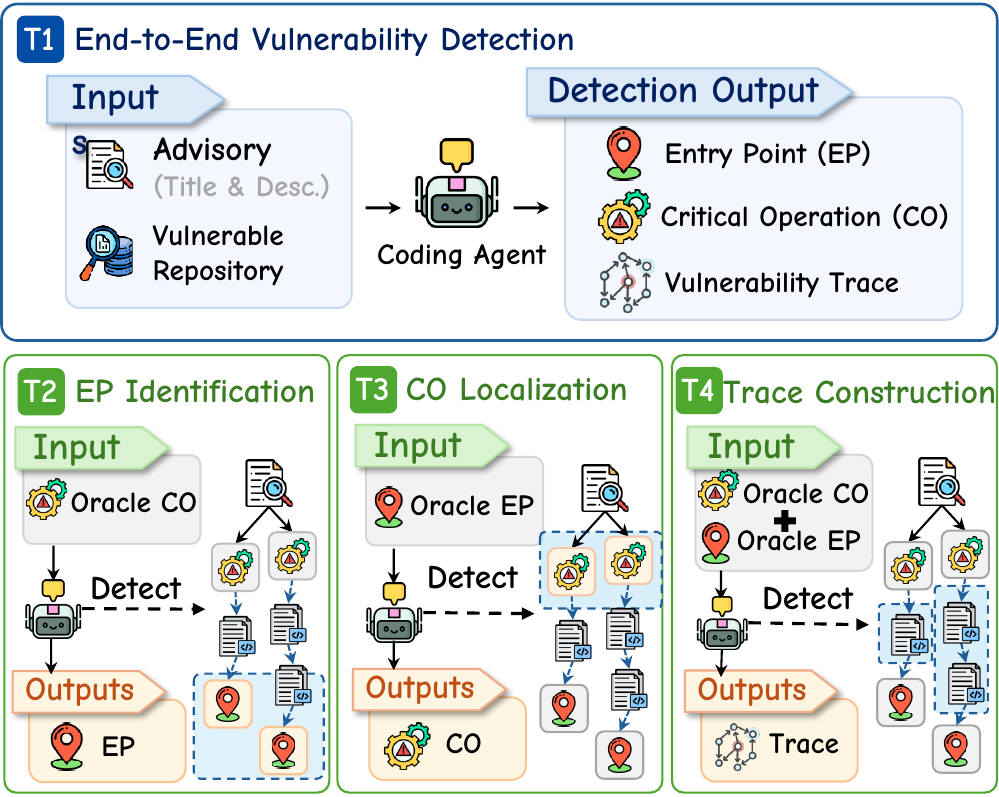}
  \caption{Overview of the four tasks in VulnGym. T1 performs end-to-end vulnerability detection from an advisory and a vulnerable repository, whereas T2--T4 provide oracle components to isolate EP identification, CO localization, and Trace construction.}
  \label{fig:Task_definition}
\end{figure}


\definecolor{clrEasy}{HTML}{009E73}
\definecolor{clrMedium}{HTML}{0072B2}
\definecolor{clrHard}{HTML}{D55E00}

\begin{figure*}[t]
\centering
\resizebox{0.95\textwidth}{!}{%
\begin{tikzpicture}
  \pgfplotsset{
    rq1a panel/.style={
      width=5.2cm, height=4.2cm,
      xtick={1,2,3,4},
      xticklabels={2B, 4B, 9B, 27B},
      xmin=0.6, xmax=4.4,
      ymin=0, ymax=0.35,
      ytick={0, 0.10, 0.20, 0.30},
      yticklabels={0\%, 10\%, 20\%, 30\%},
      tick align=inside,
      grid=both,
      major grid style={line width=0.55pt, draw=gray!25},
      tick label style={font=\footnotesize},
      major tick length=2pt,
      xtick pos=bottom,
      ytick pos=left,
      label style={font=\footnotesize},
      title={},                          
      every axis plot/.append style={line width=1.5pt, mark size=2.4pt},
    }
  }

  \begin{groupplot}[
    rq1a panel,
    group style={
      group size=3 by 1,
      horizontal sep=1.8cm,
    },
  ]

  \nextgroupplot[
      xlabel={Model Size},
    legend to name=rq1a-legend,
    legend style={
      legend columns=3,
      font=\footnotesize,
      fill=white,
      rounded corners=2pt,
      inner sep=2pt,
      column sep=6mm,
    },
  ]
  \addplot[color=clrEasy,   mark=o,
           mark options={fill=white, draw=clrEasy,   line width=1.4pt}]
    coordinates {(1,0.0091)(2,0.0818)(3,0.0636)(4,0.2818)};
  \addlegendentry{Easy}

  \addplot[color=clrMedium, mark=square,
           mark options={fill=white, draw=clrMedium, line width=1.4pt}]
    coordinates {(1,0)(2,0)(3,0.0323)(4,0.1613)};
  \addlegendentry{Medium}

  \addplot[color=clrHard,   mark=diamond,
           mark options={fill=white, draw=clrHard,   line width=1.4pt}]
    coordinates {(1,0)(2,0)(3,0)(4,0.1290)};
  \addlegendentry{Hard}

  \nextgroupplot[
    xlabel={Model Size},
    ymax=0.20,
    ytick={0, 0.05, 0.10, 0.15, 0.20},
    yticklabels={0\%, 5\%, 10\%, 15\%, 20\%},
  ]
  \addplot[color=clrEasy,   mark=o,
           mark options={fill=white, draw=clrEasy,   line width=1.4pt}]
    coordinates {(1,0.0072)(2,0.0468)(3,0.0432)(4,0.1511)};
  \addplot[color=clrMedium, mark=square,
           mark options={fill=white, draw=clrMedium, line width=1.4pt}]
    coordinates {(1,0)(2,0)(3,0.0179)(4,0.0893)};
  \addplot[color=clrHard,   mark=diamond,
           mark options={fill=white, draw=clrHard,   line width=1.4pt}]
    coordinates {(1,0)(2,0)(3,0)(4,0.0870)};

  \nextgroupplot[
    xlabel={Model Size},
    ymax=0.20,
    ytick={0, 0.05, 0.10, 0.15, 0.20},
    yticklabels={0\%, 5\%, 10\%, 15\%, 20\%},
  ]
  \addplot[color=clrEasy,   mark=o,
           mark options={fill=white, draw=clrEasy,   line width=1.4pt}]
    coordinates {(1,0)(2,0.0500)(3,0.0252)(4,0.1566)};
  \addplot[color=clrMedium, mark=square,
           mark options={fill=white, draw=clrMedium, line width=1.4pt}]
    coordinates {(1,0)(2,0)(3,0.0167)(4,0.0903)};
  \addplot[color=clrHard,   mark=diamond,
           mark options={fill=white, draw=clrHard,   line width=1.4pt}]
    coordinates {(1,0)(2,0)(3,0)(4,0.0455)};

  \end{groupplot}

  \node[anchor=south, yshift=0.15cm] at (group c2r1.north)
    {\pgfplotslegendfromname{rq1a-legend}};

  \node[anchor=north, yshift=-0.9cm, font=\small\bfseries] at (group c1r1.south)
    {(a) Advisory Recall};
  \node[anchor=north, yshift=-0.9cm, font=\small\bfseries] at (group c2r1.south)
    {(b) Entry Recall};
  \node[anchor=north, yshift=-0.9cm, font=\small\bfseries] at (group c3r1.south)
    {(c) Edit Similarity};

\end{tikzpicture}
}
\caption{Scaling performance of Qwen3.5 models under Claude Code across the Easy, Medium, and Hard subsets.}
\label{fig:rq1a_scaling}
\end{figure*}
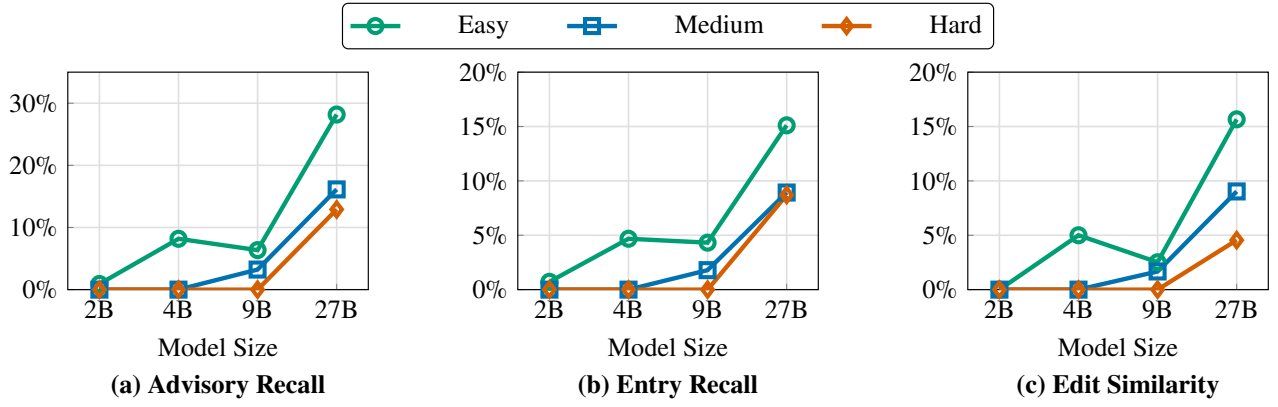

As shown in Figure~\ref{fig:Task_definition}, VulnGym defines four repository-level tasks comprising one end-to-end vulnerability detection task (T1) and three oracle-conditioned subtasks (T2--T4). 
The upper panel illustrates the end-to-end setting, while the lower panels show how oracle components isolate the capabilities required to identify EPs, localize COs, and construct vulnerability Traces.

\paragraph{End-to-End Vulnerability Detection.}
In T1, the coding agent receives a vulnerable repository together with the advisory title and description, and identifies one or more vulnerability entries comprising an EP, a CO, and their connecting Trace.
The task jointly evaluates repository exploration, vulnerability localization, reachability analysis, and evidence construction.

\paragraph{Oracle-Conditioned Tasks.}
For fine-grained capability evaluation in end-to-end vulnerability detection, we further construct three oracle-conditioned tasks.
\begin{itemize}[leftmargin=*,itemsep=1pt,topsep=2pt]
  \item \textbf{T2: Entry-Point Identification} provides an oracle CO and requires the coding agent to identify the externally reachable EPs that can lead to it.
  \item \textbf{T3: Critical-Operation Localization} provides an oracle EP and requires the coding agent to locate all COs reachable from it.
  \item \textbf{T4: Trace Construction} provides an oracle-bound EP--CO pair and requires the coding agent to construct its ordered evidence Trace between them.
\end{itemize}

\subsection{Evaluation Metrics}

Across all four tasks, evaluation is based on vulnerability entries represented as
\[
  \mathrm{EP}\rightarrow\mathrm{Trace}\rightarrow\mathrm{CO}.
\]
An advisory comprises one or more entries, each corresponding to an independently reachable EP--CO path.
We compare predicted and reference entry components using a unified line-level location-matching criterion: two code locations match when they share the same normalized repository-relative path and are within \(k=5\) lines of each other.
Accordingly, we next introduce the metrics for each task.

\paragraph{End-to-End Vulnerability Detection.}
For T1, we report Entry Recall (ER), Advisory Recall (AR), and Edit Similarity (ES).
A reference entry is hit when a prediction matches both its EP and CO locations.
\begin{itemize}[leftmargin=*,itemsep=1pt,topsep=2pt]
  \item \textbf{ER} measures the proportion of reference entries hit:
  \begin{equation}
    \mathrm{ER}=\frac{N_{\mathrm{entry}}^{\mathrm{hit}}}{N_{\mathrm{entry}}},
  \end{equation}
  where \(N_{\mathrm{entry}}\) is the number of reference entries and \(N_{\mathrm{entry}}^{\mathrm{hit}}\) the number hit.

  \item \textbf{AR} measures the proportion of reference advisories hit, where an advisory is hit if at least one of its entries is hit:
  \begin{equation}
    \mathrm{AR}=\frac{N_{\mathrm{adv}}^{\mathrm{hit}}}{N_{\mathrm{adv}}},
  \end{equation}
  where \(N_{\mathrm{adv}}\) is the number of reference advisories and \(N_{\mathrm{adv}}^{\mathrm{hit}}\) the number hit.

  \item \textbf{ES} measures code-location and invocation-order similarity between predicted and reference Traces for hit entries:
  \begin{equation}
    \mathrm{ES}(P,G)
    =1-\frac{D_{\mathrm{Lev}}(P,G)}{\max(n,m)},
    \label{eq:edit_similarity}
  \end{equation}
  where \(P=(p_1,\ldots,p_n)\) and \(G=(g_1,\ldots,g_m)\) are the predicted and reference Traces, and \(D_{\mathrm{Lev}}\) is their Levenshtein edit distance.
\end{itemize}

\paragraph{EP Identification and CO Localization.}
The EP-identification and CO-localization tasks share three metrics: File Recall (FR), Line Recall (LR), and Finding Precision (FP).
Given oracle COs, T2 evaluates predicted EPs, while T3 evaluates predicted COs under oracle EPs.
\begin{itemize}[leftmargin=*,itemsep=1pt,topsep=2pt]
  \item \textbf{FR} measures the proportion of reference targets localized to the correct file:
  \begin{equation}
    \mathrm{FR}=\frac{M_{\mathrm{file}}}{N_G},
  \end{equation}
  where \(M_{\mathrm{file}}\) is the number of file-level matches and \(N_G\) the number of reference targets.

  \item \textbf{LR} measures the percentage of reference targets matched by a prediction within \(k\) lines:
  \begin{equation}
    \mathrm{LR}=\frac{M_k}{N_G},
  \end{equation}
  where \(M_k\) is the number of line-level matches.

  \item \textbf{FP} measures the percentage of predictions that correctly match a reference target within \(k\) lines:
  \begin{equation}
    \mathrm{FP}=\frac{M_k}{N_P},
  \end{equation}
  where \(N_P\) is the number of predictions.
\end{itemize}

\paragraph{Trace Construction.}
T4 reuses ES and additionally reports Trace Node Recall (TNR) and Trace Node Precision (TNP).
\begin{itemize}[leftmargin=*,itemsep=1pt,topsep=2pt]
  \item \textbf{TNR} measures the percentage of reference intermediate nodes matched by a prediction:
  \begin{equation}
    \mathrm{TNR}=\frac{M_I}{N_I^G},
  \end{equation}
  where \(M_I\) is the number of one-to-one node matches and \(N_I^G\) the number of reference intermediate nodes.

  \item \textbf{TNP} measures the percentage of predicted intermediate nodes that match a reference node:
  \begin{equation}
    \mathrm{TNP}=\frac{M_I}{N_I^P},
  \end{equation}
  where \(N_I^P\) is the number of predicted intermediate nodes.
\end{itemize}

\section{Experiment} \label{sec:Exp}
    \subsection{Experimental Setup}
\subsubsection{LLM Selection}
To support repository-level code exploration, we select models with context windows of at least 256K tokens  
Specifically, we use Qwen3.5-2B, 4B, 9B, and 27B~\cite{qwen2026qwen35} for scaling analysis within a single model family. 
We also include three frontier LLMs for model comparison: DeepSeek-V4-Flash~\cite{xu2026deepseek}, MiniMax-M3~\cite{minimax2026m3}, and GLM-5.2~\cite{zai2026glm52}.

\begin{table}[t]
  \centering
  {\small
  \setlength{\tabcolsep}{1mm}
  \begin{threeparttable}
    \begin{tabular}{ll rrr}
      \toprule
      \multirow{2}{*}{\textbf{Scaffold}}
        & \multirow{2}{*}{\textbf{LLM}}
        & \multicolumn{3}{c}{\textbf{End-to-End Detection}} \\
      \cmidrule(lr){3-5}
        & & \textbf{AR} (\%)$\uparrow$ & \textbf{ER} (\%)$\uparrow$ & \textbf{ES} (\%)$\uparrow$ \\
      \midrule
      \multirow{3}{*}{Claude Code}
        & DeepSeek-V4-Flash & 9.68  & 6.52  & 3.37 \\
        & GLM-5.2           & \textbf{12.90} & \textbf{8.70} & \textbf{6.94} \\
        & MiniMax-M3        & \textbf{12.90} & \textbf{8.70} & 6.25 \\
      \midrule
      \multirow{3}{*}{OpenHands}
        & DeepSeek-V4-Flash & \textbf{22.58} & \textbf{15.22} & \textbf{11.63} \\
        & GLM-5.2           & 6.45  & 4.35  & 2.15 \\
        & MiniMax-M3        & 16.13 & 10.87 & 7.93 \\
      \midrule
      \multirow{3}{*}{MiniSWE}
        & DeepSeek-V4-Flash & \textbf{16.13} & \textbf{10.87} & 7.67 \\
        & GLM-5.2           & \textbf{16.13} & \textbf{10.87} & \textbf{8.65} \\
        & MiniMax-M3        & \textbf{16.13} & \textbf{10.87} & 6.95 \\
      \bottomrule
    \end{tabular}
  \end{threeparttable}
  }
  \caption{%
    End-to-end vulnerability performance of frontier coding agent on Hard cases.
  }
  \label{tab:rq2_frontier}
\end{table}

\subsubsection{Agentic Scaffolds}
We evaluate three widely used coding-agent scaffolds: Claude Code~\cite{anthropic2025claudecode}, OpenHands~\cite{openhands2024}, and MiniSWE~\cite{yang2024sweagent}.
For each task, the agent is restricted to invoking only read-only commands, including \texttt{cat}, \texttt{find}, \texttt{grep}, \texttt{head}, \texttt{ls}, \texttt{nl}, \texttt{pwd}, \texttt{rg}, \texttt{sed}, \texttt{tail}, and \texttt{wc}.

\subsection{Model Scaling across Difficulty Levels}
We stratify VulnGym's verified entries by Trace length into Easy ($\leq 7$), Medium ($=8$), and Hard ($\geq 9$), keeping Medium and Hard comparable, comprising 110/31/31 advisories and 278/56/46 entries, respectively.
To investigate model scaling, we evaluate Qwen3.5-2B, 4B, 9B, and 27B using Claude Code as the agentic scaffold and report their T1 performance at each difficulty level.
\textbf{With model scaling, coding agents generally achieve stronger end-to-end vulnerability detection.}
As shown in Figure~\ref{fig:rq1a_scaling}, the 2B and 4B models obtain non-zero scores only on Easy, while 9B extends non-zero performance to Medium.
Qwen3.5-27B leads on every metric at every difficulty level and is the only model to achieve non-zero performance on Hard.
Notably, scaling up model size does not consistently lead to better performance.
4B outperforms 9B on all three metrics at the Easy level, suggesting that the gains from model scaling may vary with case difficulty.

\textbf{Across model sizes, end-to-end vulnerability detection performance declines consistently as case difficulty increases.}
For Qwen3.5-9B, AR, ER, and ES decrease from 6.36\%, 4.32\%, and 2.52\% on Easy to 3.23\%, 1.79\%, and 1.67\% on Medium, respectively, and all reach zero on Hard.
The 2B and 4B variants reach zero on all three metrics at Medium and remain on Hard.
Even Qwen3.5-27B achieves merely 12.90\% AR, 8.70\% ER, and 4.55\% ES on Hard.

\subsection{Frontier Coding Agents on Hard Cases}
Although Qwen3.5-27B performs best among the scaling models, its end-to-end performance on Hard remains limited.
To assess whether stronger coding agents can improve upon these limited results, we pair DeepSeek-V4-Flash, GLM-5.2, and MiniMax-M3 with all three scaffolds and evaluate the resulting nine frontier coding agents on T1 over Hard level.

\textbf{Frontier coding agents improve over Qwen3.5-27B, but Hard entries remain largely unresolved.}
As shown in Table~\ref{tab:rq2_frontier}, DeepSeek-V4-Flash with OpenHands performs best, achieving 22.58\% AR, 15.22\% ER, and 11.63\% ES, compared with 12.90\%, 8.70\%, and 4.55\% for Qwen3.5-27B with Claude Code.
Nevertheless, even this best-performing coding agent detects fewer than one quarter of hard-level entries, leaving substantial room for improvement.

\textbf{Coding-agent performance varies across LLM--scaffold combinations, with scaffolds differing in their sensitivity to LLM selection.}
DeepSeek-V4-Flash performs best with OpenHands, but its performance drops to 9.68\% AR, 6.52\% ER, and 3.37\% ES with Claude Code.
By contrast, GLM-5.2 exhibits the opposite pattern, with performance increasing from 6.45\% AR, 4.35\% ER, and 2.15\% ES with OpenHands to 12.90\%, 8.70\%, and 6.94\% with Claude Code.
Unlike these model-specific reversals, MiniSWE yields the same AR of 16.13\% and ER of 10.87\% across all three LLMs, while its ES ranges narrowly from 6.95\% to 8.65\%.
These results indicate coding agent compatibility is critical to end-to-end performance, while showing that MiniSWE remains comparatively stable across the evaluated LLMs.

\definecolor{clrNoUsable}{HTML}{E15759}     
\definecolor{clrNoEndpoint}{HTML}{F28E2B}   
\definecolor{clrEndpointOnly}{HTML}{4E79A7} 
\definecolor{clrEPMiss}{HTML}{59A14F}       
\definecolor{clrCOMiss}{HTML}{BAB0AC}       

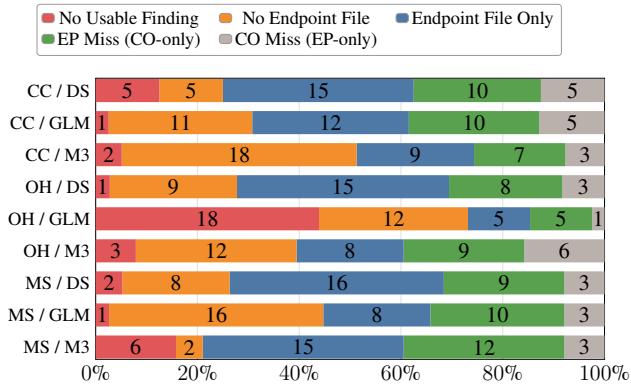
\begin{figure}[t]
\centering
\resizebox{\columnwidth}{!}{%
\begin{tikzpicture}

\begin{axis}[
    width=13.2cm, height=8.0cm,
    xbar stacked,
    xmin=0, xmax=100,
    xtick={0,20,40,60,80,100},
    xticklabel={\pgfmathprintnumber{\tick}\%},
    tick align=inside,
    grid=major,
    major grid style={line width=0.4pt, draw=gray!25},
    tick label style={font=\LARGE},
    label style={font=\LARGE},
    major tick length=2pt,
    xtick pos=bottom,
    ytick pos=left,
    bar width=15pt,
    every axis plot/.append style={draw=none},
    point meta=explicit symbolic,
    nodes near coords={\pgfplotspointmeta},
    every node near coord/.append style={
      font=\LARGE,
      text=black,
      anchor=center,
      inner sep=0pt,
    },
    yticklabel style={font=\Large},
    ytick={1,...,9},
    yticklabels={CC / DS, CC / GLM, CC / M3,
                 OH / DS, OH / GLM, OH / M3,
                 MS / DS, MS / GLM, MS / M3},
    y dir=reverse,
    enlarge y limits={abs=8pt},
    legend style={
      at={(0.40,1.08)},
      anchor=south,
      draw=black!40,
      fill=white,
      rounded corners=2pt,
      inner sep=3pt,
      font=\Large,
      /tikz/every even column/.append style={column sep=0.35cm},
    },
    legend columns=3,
  ]

  \addplot[fill=clrNoUsable] coordinates {
    (12.5000,1) [5] (2.5641,2) [1] (5.1282,3) [2] (2.7778,4) [1] (43.9024,5) [18]
    (7.8947,6) [3] (5.2632,7) [2] (2.6316,8) [1] (15.7895,9) [6]
  };
  \addlegendentry{No Usable Finding}
  \addplot[fill=clrNoEndpoint] coordinates {
    (12.5000,1) [5] (28.2051,2) [11] (46.1538,3) [18] (25.0000,4) [9] (29.2683,5) [12]
    (31.5789,6) [12] (21.0526,7) [8] (42.1053,8) [16] (5.2632,9) [2]
  };
  \addlegendentry{No Endpoint File}
  \addplot[fill=clrEndpointOnly] coordinates {
    (37.5000,1) [15] (30.7692,2) [12] (23.0769,3) [9] (41.6667,4) [15] (12.1951,5) [5]
    (21.0526,6) [8] (42.1053,7) [16] (21.0526,8) [8] (39.4737,9) [15]
  };
  \addlegendentry{Endpoint File Only}
  \addplot[fill=clrEPMiss] coordinates {
    (25.0000,1) [10] (25.6410,2) [10] (17.9487,3) [7] (22.2222,4) [8] (12.1951,5) [5]
    (23.6842,6) [9] (23.6842,7) [9] (26.3158,8) [10] (31.5789,9) [12]
  };
  \addlegendentry{EP Miss (CO-only)}
  \addplot[fill=clrCOMiss] coordinates {
    (12.5000,1) [5] (12.8205,2) [5] (7.6923,3) [3] (8.3333,4) [3] (2.4390,5) [1]
    (15.7895,6) [6] (7.8947,7) [3] (7.8947,8) [3] (7.8947,9) [3]
  };
  \addlegendentry{CO Miss (EP-only)}

\end{axis}

\end{tikzpicture}
}
\caption{Composition of relation-level T1 failures on Hard. CC: Claude Code; OH: OpenHands; MS: MiniSWE; DS: DeepSeek-V4-Flash; GLM: GLM-5.2; M3: MiniMax-M3.}
\label{fig:t1_failure_composition}
\end{figure}

\textbf{T1 failures mainly involve missing the vulnerable files entirely or failing at line-level localization.}
Figure~\ref{fig:t1_failure_composition} summarizes the composition of 347 relation-level T1 failures.
No Endpoint File indicates that neither predicted EP or CO matches its corresponding reference file (93; 26.8\%), whereas Endpoint File Only means that at least one endpoint reaches the correct file but neither is localized to the correct line (103; 29.7\%).
Together, these two categories account for 56.5\% of all failures.
When one endpoint is localized, EP Miss (CO-only) is substantially more common than CO Miss (EP-only), with 80 (23.1\%) versus 32 (9.2\%) cases, suggesting that EP identification is more frequently unresolved in T1 outputs.
No Usable Finding accounts for the remaining 39 failures (11.2\%), including 18 from OpenHands with GLM-5.2 alone.
Overall, these results reveal where end-to-end predictions fall short, but provide only a coarse-grained view of the underlying capability limitations.

\subsection{Oracle-Conditioned Capability Analysis}
To better understand these limitations, we next perform oracle-conditioned evaluation to analyze the capabilities underlying end-to-end vulnerability detection.

\begin{table*}[t]
  \centering
  \footnotesize
  \setlength{\tabcolsep}{4pt}
  \begin{threeparttable}
    \begin{tabular}{@{}ll | rrr | rrr | rrr@{}}
      \toprule
      \multirow{2}{*}{\textbf{Scaffold}}
        & \multirow{2}{*}{\textbf{LLM}}
        & \multicolumn{3}{c|}{\textbf{EP Identification}}
        & \multicolumn{3}{c|}{\textbf{CO Localization}}
        & \multicolumn{3}{c}{\textbf{Trace Construction}} \\
      \cmidrule(lr){3-5}\cmidrule(lr){6-8}\cmidrule(lr){9-11}
        & & \textbf{FR} (\%)$\uparrow$ & \textbf{LR} (\%)$\uparrow$ & \textbf{FP} (\%)$\uparrow$
        & \textbf{FR} (\%)$\uparrow$ & \textbf{LR} (\%)$\uparrow$ & \textbf{FP} (\%)$\uparrow$
        & \textbf{ES} (\%)$\uparrow$ & \textbf{TNR} (\%)$\uparrow$ & \textbf{TNP} (\%)$\uparrow$ \\
      \midrule
      \multirow{3}{*}{Claude Code}
        & DeepSeek-V4-Flash & 13.95 & 4.65 & 1.69 & 34.88 & 16.28 & 0.91 & \textbf{34.68} & \textbf{53.96} & \textbf{20.22} \\
        & GLM-5.2           & \textbf{25.58} & \textbf{9.30} & \textbf{5.97} & \textbf{48.84} & \textbf{30.23} & \textbf{5.22} & 32.07 & 39.86 & 19.79 \\
        & MiniMax-M3        & 13.95 & 2.33 & 0.79 & 39.53 & 16.28 & 1.23 & 22.01 & 53.60 & 14.61 \\
      \midrule
      \multirow{3}{*}{OpenHands}
        & DeepSeek-V4-Flash & 18.60 & 2.33 & 1.01 & 37.21 & 20.93 & 2.23 & 26.14 & 41.64 & 19.63 \\
        & GLM-5.2           & 18.60 & 9.30 & \textbf{8.16} & 6.98 & 2.33 & 1.33 & 16.47 & 30.60 & \textbf{26.71} \\
        & MiniMax-M3        & \textbf{23.26} & \textbf{16.28} & 8.05 & \textbf{46.51} & \textbf{23.26} & \textbf{3.25} & \textbf{26.37} & \textbf{52.67} & 19.53 \\
      \midrule
      \multirow{3}{*}{MiniSWE}
        & DeepSeek-V4-Flash & 16.28 & 6.98 & 3.49 & \textbf{51.16} & 16.28 & 1.64 & 27.77 & 37.72 & 16.13 \\
        & GLM-5.2           & \textbf{30.23} & \textbf{18.60} & \textbf{11.11} & 39.53 & \textbf{18.60} & 2.54 & \textbf{36.37} & \textbf{49.47} & \textbf{23.28} \\
        & MiniMax-M3        & 13.95 & 6.98 & 4.69 & 46.51 & \textbf{18.60} & \textbf{3.11} & 30.20 & 39.86 & 20.07 \\
      \bottomrule
    \end{tabular}
  \end{threeparttable}
  \caption{%
    Oracle-conditioned tasks performance on Hard cases.
    Within each scaffold, bold values denote the best performance in each column.
  }
  \label{tab:rq3_diagnosis}
\end{table*}

\textbf{Under the respective oracle conditions, EP identification is generally more challenging than CO localization.}
As shown in Table~\ref{tab:rq3_diagnosis}, CO localization yields higher FR across all LLM--scaffold combinations except OpenHands with GLM-5.2.
For LR, only OpenHands with GLM-5.2 yields higher performance on EP identification than on CO localization among the nine coding agents, reaching 9.30\% and 2.33\%, respectively.
Across all coding agents, the highest CO-localization FR and LR are 51.16\% and 30.23\%, respectively, compared with 30.23\% and 18.60\% for EP identification.
However, FP reveals a precision--recall trade-off, as EP identification generally achieves higher scores than CO localization, indicating more selective but lower-coverage predictions.
Nevertheless, FP peaks at only 11.11\% for EP identification and 5.22\% for CO localization, indicating that most predicted findings remain unmatched at the line level.

\textbf{Even with the oracle EP--CO pair provided, the vulnerable evidence Traces remain incomplete and noisy.}
Across coding agents, TNR ranges from 30.60\% to 53.96\%, whereas TNP remains substantially lower, ranging from 14.61\% to 26.71\%.
This recall--precision imbalance indicates that coding agents recover some reference intermediate nodes but include many unmatched nodes in their predictions.
After endpoint uncertainty is removed, the highest ES is still only 36.37\%, indicating persistent difficulty in constructing accurate evidence Traces.

\section{Discussion} \label{sec:discussion}
    \subsection{Practical Efficiency Analysis}

\definecolor{clrExplore}{HTML}{59A14F}
\definecolor{clrSearch}{HTML}{4E79A7}
\definecolor{clrRead}{HTML}{F28E2B}
\definecolor{clrOther}{HTML}{BAB0AC}
\definecolor{clrFailed}{HTML}{E15759}
\definecolor{clrDSFlash}{HTML}{2166AC}
\definecolor{clrGLM}{HTML}{B2182B}
\definecolor{clrMiniMax}{HTML}{1B7837}
\definecolor{clrLegendGray}{HTML}{999999}
\definecolor{clrTierLow}{HTML}{CFE8F7}
\definecolor{clrTierMid}{HTML}{DCEEF8}
\definecolor{clrTierHigh}{HTML}{ECF5FA}

\begin{figure*}[t]
\centering
\resizebox{0.95\textwidth}{!}{%
\begin{tikzpicture}

  \pgfplotsset{
    info panel/.style={
      width=5.2cm, height=4.2cm,
      ybar stacked,
      ymin=0,
      tick align=inside,
      grid=both,
      major grid style={line width=0.4pt, draw=gray!25},
      tick label style={font=\tiny},
      label style={font=\footnotesize},
      xlabel style={yshift=2pt},
      title style={font=\footnotesize\bfseries, yshift=-2pt},
      major tick length=2pt,
      xtick pos=bottom,
      ytick pos=left,
      every axis plot/.append style={draw=none},
    }
  }

  \begin{groupplot}[
    info panel,
    group style={
      group size=3 by 1,
      horizontal sep=1.4cm,
      vertical sep=0.9cm,
    },
  ]

  \nextgroupplot[
      title={Claude Code},
      ylabel={Number of Actions},
      xlabel={Turns},
      xtick={0,20,40,60,80,100},
      xmin=-1, xmax=111,
      ymax=60,
      ytick={0,10,20,30,40,50,60},
      bar width=1.3pt,
      legend to name=infolegend,
      legend style={legend columns=5, font=\footnotesize, column sep=5mm, draw=none},
  ]
  \addplot[fill=clrExplore] coordinates {
    (1,32) (2,34) (3,23) (4,18) (5,16) (6,10) (7,16) (8,15)
    (9,15) (10,6) (11,5) (12,3) (13,2) (14,4) (15,4) (16,1)
    (17,1) (18,1) (19,4) (20,2) (21,7) (22,4) (23,4) (24,0)
    (25,3) (26,1) (27,2) (28,0) (29,1) (30,1) (31,1) (32,0)
    (33,0) (34,0) (35,0) (36,1) (37,2) (38,3) (39,1) (40,0)
    (41,1) (42,2) (43,1) (44,1) (45,0) (46,0) (47,1) (48,0)
    (49,1) (50,1) (51,1) (52,0) (53,0) (54,0) (55,0) (56,0)
    (57,0) (58,0) (59,0) (60,0) (61,0) (62,0) (63,0) (64,0)
    (65,0) (66,1) (67,0) (68,0) (69,2) (70,0) (71,0) (72,0)
    (73,0) (74,0) (75,0) (76,0) (77,1) (78,0) (79,0) (80,0)
    (81,0) (82,0) (83,0) (84,0) (85,0) (86,0) (87,0) (88,0)
    (89,0) (90,0) (91,0) (92,0) (93,0) (94,0) (95,0) (96,1)
    (97,0) (98,0) (99,0) (100,0) (101,0) (102,0) (103,0) (104,0)
    (105,0) (106,0) (107,0) (108,0) (109,0)
  };
  \addlegendentry{Explore}
  \addplot[fill=clrSearch] coordinates {
    (1,0) (2,1) (3,4) (4,7) (5,7) (6,5) (7,5) (8,7)
    (9,6) (10,5) (11,1) (12,5) (13,6) (14,4) (15,8) (16,8)
    (17,12) (18,5) (19,7) (20,12) (21,7) (22,9) (23,10) (24,12)
    (25,8) (26,12) (27,11) (28,8) (29,7) (30,12) (31,9) (32,9)
    (33,14) (34,17) (35,17) (36,9) (37,11) (38,14) (39,9) (40,11)
    (41,8) (42,13) (43,10) (44,14) (45,13) (46,13) (47,13) (48,10)
    (49,12) (50,12) (51,14) (52,10) (53,9) (54,12) (55,8) (56,11)
    (57,5) (58,8) (59,7) (60,12) (61,10) (62,9) (63,3) (64,5)
    (65,4) (66,4) (67,7) (68,2) (69,4) (70,2) (71,4) (72,6)
    (73,3) (74,3) (75,3) (76,1) (77,2) (78,7) (79,3) (80,4)
    (81,3) (82,1) (83,6) (84,3) (85,6) (86,2) (87,2) (88,1)
    (89,1) (90,2) (91,1) (92,0) (93,3) (94,0) (95,2) (96,0)
    (97,0) (98,0) (99,1) (100,1) (101,2) (102,0) (103,0) (104,0)
    (105,0) (106,0) (107,0) (108,0) (109,0)
  };
  \addlegendentry{Search}
  \addplot[fill=clrRead] coordinates {
    (1,26) (2,6) (3,13) (4,10) (5,14) (6,13) (7,13) (8,13)
    (9,15) (10,23) (11,22) (12,23) (13,20) (14,20) (15,27) (16,23)
    (17,24) (18,24) (19,21) (20,20) (21,20) (22,21) (23,17) (24,19)
    (25,22) (26,20) (27,21) (28,25) (29,22) (30,24) (31,20) (32,25)
    (33,21) (34,20) (35,19) (36,20) (37,18) (38,17) (39,22) (40,19)
    (41,21) (42,19) (43,25) (44,18) (45,22) (46,19) (47,18) (48,19)
    (49,15) (50,16) (51,19) (52,15) (53,15) (54,14) (55,16) (56,11)
    (57,18) (58,13) (59,14) (60,11) (61,10) (62,11) (63,13) (64,12)
    (65,11) (66,9) (67,8) (68,8) (69,6) (70,10) (71,7) (72,6)
    (73,5) (74,6) (75,8) (76,6) (77,6) (78,6) (79,6) (80,7)
    (81,7) (82,4) (83,3) (84,4) (85,1) (86,2) (87,1) (88,3)
    (89,3) (90,1) (91,2) (92,3) (93,3) (94,3) (95,2) (96,2)
    (97,3) (98,3) (99,1) (100,1) (101,0) (102,1) (103,1) (104,1)
    (105,1) (106,1) (107,1) (108,1) (109,1)
  };
  \addlegendentry{Read}
  \addplot[fill=clrOther] coordinates {
    (1,1) (2,8) (3,6) (4,10) (5,5) (6,2) (7,4) (8,6)
    (9,9) (10,5) (11,7) (12,5) (13,7) (14,9) (15,5) (16,1)
    (17,0) (18,0) (19,2) (20,4) (21,5) (22,3) (23,0) (24,0)
    (25,2) (26,1) (27,4) (28,0) (29,3) (30,2) (31,3) (32,3)
    (33,0) (34,0) (35,0) (36,4) (37,2) (38,6) (39,1) (40,0)
    (41,0) (42,1) (43,1) (44,1) (45,1) (46,0) (47,3) (48,3)
    (49,3) (50,1) (51,0) (52,0) (53,0) (54,1) (55,2) (56,1)
    (57,0) (58,0) (59,0) (60,5) (61,0) (62,0) (63,2) (64,0)
    (65,3) (66,0) (67,1) (68,1) (69,3) (70,2) (71,0) (72,0)
    (73,0) (74,1) (75,0) (76,0) (77,0) (78,0) (79,0) (80,0)
    (81,0) (82,0) (83,0) (84,0) (85,0) (86,2) (87,1) (88,0)
    (89,0) (90,0) (91,0) (92,0) (93,0) (94,0) (95,0) (96,0)
    (97,0) (98,0) (99,0) (100,0) (101,0) (102,1) (103,1) (104,1)
    (105,0) (106,0) (107,0) (108,0) (109,0)
  };
  \addlegendentry{Other}
  \addplot[fill=clrFailed] coordinates {
    (1,0) (2,0) (3,2) (4,2) (5,5) (6,8) (7,10) (8,4)
    (9,1) (10,1) (11,3) (12,2) (13,2) (14,0) (15,0) (16,1)
    (17,1) (18,0) (19,0) (20,0) (21,1) (22,0) (23,0) (24,2)
    (25,0) (26,1) (27,0) (28,0) (29,0) (30,0) (31,0) (32,0)
    (33,0) (34,0) (35,0) (36,0) (37,1) (38,1) (39,1) (40,0)
    (41,0) (42,0) (43,0) (44,1) (45,0) (46,0) (47,1) (48,0)
    (49,0) (50,0) (51,0) (52,0) (53,0) (54,0) (55,0) (56,0)
    (57,0) (58,1) (59,0) (60,0) (61,0) (62,0) (63,0) (64,1)
    (65,0) (66,0) (67,0) (68,0) (69,0) (70,0) (71,0) (72,0)
    (73,0) (74,0) (75,0) (76,0) (77,0) (78,0) (79,0) (80,0)
    (81,0) (82,0) (83,0) (84,0) (85,0) (86,0) (87,0) (88,0)
    (89,0) (90,0) (91,0) (92,0) (93,0) (94,0) (95,0) (96,0)
    (97,0) (98,0) (99,0) (100,0) (101,0) (102,0) (103,0) (104,0)
    (105,0) (106,0) (107,0) (108,0) (109,0)
  };
  \addlegendentry{Failed}

  \nextgroupplot[
      title={MiniSWE},
      yticklabel={},
      xlabel={Turns},
      xtick={0,20,40,60,80,100},
      xmin=-1, xmax=118,
      ymax=70,
      ytick={0,10,20,30,40,50,60,70},
      bar width=1.2pt,
  ]
  \addplot[fill=clrExplore] coordinates {
    (1,22) (2,24) (3,19) (4,16) (5,11) (6,13) (7,8) (8,11)
    (9,5) (10,5) (11,6) (12,6) (13,5) (14,5) (15,3) (16,5)
    (17,5) (18,5) (19,4) (20,2) (21,2) (22,1) (23,2) (24,3)
    (25,3) (26,1) (27,3) (28,1) (29,4) (30,1) (31,3) (32,2)
    (33,2) (34,3) (35,1) (36,0) (37,0) (38,2) (39,1) (40,1)
    (41,1) (42,1) (43,2) (44,0) (45,1) (46,0) (47,0) (48,3)
    (49,1) (50,1) (51,2) (52,1) (53,1) (54,0) (55,0) (56,0)
    (57,0) (58,0) (59,0) (60,0) (61,0) (62,0) (63,0) (64,0)
    (65,0) (66,0) (67,1) (68,0) (69,1) (70,0) (71,0) (72,0)
    (73,0) (74,0) (75,0) (76,0) (77,0) (78,0) (79,0) (80,1)
    (81,0) (82,0) (83,1) (84,2) (85,0) (86,1) (87,0) (88,1)
    (89,1) (90,0) (91,0) (92,0) (93,0) (94,0) (95,0) (96,0)
    (97,0) (98,0) (99,0) (100,0) (101,0) (102,0) (103,0) (104,0)
    (105,0) (106,0) (107,0) (108,0) (109,0) (110,0) (111,0) (112,0)
    (113,0) (114,0) (115,0) (116,0)
  };
  \addplot[fill=clrSearch] coordinates {
    (1,0) (2,14) (3,17) (4,7) (5,9) (6,6) (7,7) (8,8)
    (9,14) (10,16) (11,17) (12,17) (13,6) (14,14) (15,11) (16,2)
    (17,10) (18,14) (19,11) (20,7) (21,12) (22,5) (23,13) (24,16)
    (25,10) (26,4) (27,13) (28,15) (29,11) (30,10) (31,12) (32,13)
    (33,14) (34,15) (35,12) (36,12) (37,16) (38,15) (39,17) (40,24)
    (41,14) (42,14) (43,20) (44,21) (45,14) (46,15) (47,8) (48,15)
    (49,13) (50,8) (51,8) (52,10) (53,7) (54,10) (55,10) (56,14)
    (57,13) (58,11) (59,6) (60,7) (61,9) (62,13) (63,6) (64,7)
    (65,15) (66,10) (67,6) (68,5) (69,9) (70,6) (71,6) (72,9)
    (73,1) (74,15) (75,0) (76,7) (77,6) (78,4) (79,5) (80,1)
    (81,7) (82,2) (83,3) (84,0) (85,0) (86,1) (87,0) (88,0)
    (89,4) (90,2) (91,3) (92,3) (93,0) (94,5) (95,1) (96,2)
    (97,1) (98,1) (99,1) (100,2) (101,2) (102,1) (103,1) (104,1)
    (105,6) (106,6) (107,0) (108,0) (109,1) (110,2) (111,0) (112,6)
    (113,2) (114,0) (115,0) (116,0)
  };
  \addplot[fill=clrRead] coordinates {
    (1,22) (2,12) (3,21) (4,25) (5,28) (6,28) (7,27) (8,23)
    (9,30) (10,29) (11,23) (12,26) (13,36) (14,29) (15,31) (16,38)
    (17,29) (18,28) (19,26) (20,29) (21,27) (22,33) (23,25) (24,31)
    (25,27) (26,29) (27,28) (28,27) (29,27) (30,26) (31,26) (32,26)
    (33,25) (34,24) (35,28) (36,26) (37,26) (38,20) (39,21) (40,21)
    (41,25) (42,22) (43,25) (44,24) (45,19) (46,21) (47,23) (48,19)
    (49,14) (50,19) (51,19) (52,15) (53,17) (54,18) (55,17) (56,20)
    (57,17) (58,19) (59,16) (60,18) (61,20) (62,17) (63,17) (64,18)
    (65,17) (66,16) (67,12) (68,14) (69,9) (70,12) (71,13) (72,11)
    (73,13) (74,12) (75,14) (76,12) (77,11) (78,10) (79,6) (80,8)
    (81,6) (82,8) (83,5) (84,6) (85,6) (86,3) (87,3) (88,4)
    (89,1) (90,3) (91,2) (92,0) (93,3) (94,2) (95,3) (96,3)
    (97,3) (98,3) (99,2) (100,2) (101,1) (102,2) (103,2) (104,2)
    (105,1) (106,2) (107,2) (108,1) (109,5) (110,3) (111,1) (112,0)
    (113,2) (114,1) (115,0) (116,0)
  };
  \addplot[fill=clrOther] coordinates {
    (1,0) (2,10) (3,7) (4,13) (5,12) (6,11) (7,14) (8,17)
    (9,13) (10,14) (11,15) (12,18) (13,12) (14,15) (15,13) (16,20)
    (17,19) (18,13) (19,13) (20,13) (21,14) (22,14) (23,13) (24,16)
    (25,11) (26,12) (27,12) (28,12) (29,13) (30,12) (31,19) (32,14)
    (33,13) (34,19) (35,12) (36,15) (37,14) (38,10) (39,14) (40,11)
    (41,13) (42,12) (43,11) (44,10) (45,11) (46,13) (47,14) (48,13)
    (49,13) (50,10) (51,10) (52,11) (53,10) (54,10) (55,11) (56,10)
    (57,10) (58,10) (59,11) (60,9) (61,18) (62,12) (63,11) (64,12)
    (65,17) (66,14) (67,11) (68,11) (69,11) (70,8) (71,8) (72,8)
    (73,7) (74,7) (75,7) (76,7) (77,8) (78,7) (79,8) (80,5)
    (81,5) (82,5) (83,6) (84,3) (85,3) (86,2) (87,1) (88,2)
    (89,4) (90,2) (91,1) (92,1) (93,1) (94,1) (95,1) (96,1)
    (97,1) (98,1) (99,2) (100,1) (101,1) (102,1) (103,1) (104,1)
    (105,1) (106,2) (107,1) (108,7) (109,5) (110,2) (111,2) (112,6)
    (113,2) (114,1) (115,1) (116,1)
  };
  \addplot[fill=clrFailed] coordinates {
    (1,8) (2,6) (3,3) (4,0) (5,1) (6,1) (7,2) (8,2)
    (9,0) (10,1) (11,2) (12,1) (13,0) (14,1) (15,1) (16,1)
    (17,1) (18,0) (19,0) (20,0) (21,1) (22,0) (23,1) (24,0)
    (25,0) (26,1) (27,1) (28,0) (29,0) (30,3) (31,0) (32,1)
    (33,2) (34,1) (35,1) (36,0) (37,0) (38,1) (39,1) (40,0)
    (41,0) (42,0) (43,0) (44,1) (45,2) (46,1) (47,1) (48,1)
    (49,0) (50,1) (51,0) (52,1) (53,1) (54,0) (55,0) (56,0)
    (57,1) (58,1) (59,1) (60,0) (61,0) (62,1) (63,1) (64,1)
    (65,0) (66,1) (67,3) (68,1) (69,0) (70,1) (71,0) (72,1)
    (73,1) (74,0) (75,0) (76,0) (77,0) (78,0) (79,1) (80,0)
    (81,0) (82,1) (83,1) (84,1) (85,1) (86,1) (87,2) (88,0)
    (89,0) (90,0) (91,0) (92,0) (93,0) (94,0) (95,0) (96,0)
    (97,0) (98,0) (99,0) (100,0) (101,0) (102,0) (103,0) (104,0)
    (105,0) (106,0) (107,0) (108,0) (109,0) (110,0) (111,0) (112,0)
    (113,0) (114,0) (115,0) (116,0)
  };

  \nextgroupplot[
      title={OpenHands},
      yticklabel={},
      xlabel={Turns},
      xtick={0,20,40,60,80,100,120,140},
      xmin=-1, xmax=145,
      ymax=60,
      ytick={0,10,20,30,40,50,60},
      bar width=1.0pt,
  ]
  \addplot[fill=clrExplore] coordinates {
    (1,4) (2,8) (3,5) (4,10) (5,7) (6,4) (7,3) (8,3)
    (9,3) (10,3) (11,3) (12,0) (13,0) (14,1) (15,2) (16,3)
    (17,2) (18,2) (19,2) (20,2) (21,2) (22,0) (23,0) (24,3)
    (25,1) (26,3) (27,1) (28,0) (29,3) (30,1) (31,1) (32,2)
    (33,0) (34,0) (35,2) (36,1) (37,0) (38,0) (39,1) (40,3)
    (41,0) (42,0) (43,1) (44,2) (45,0) (46,2) (47,1) (48,0)
    (49,1) (50,1) (51,0) (52,2) (53,1) (54,0) (55,0) (56,1)
    (57,0) (58,1) (59,0) (60,0) (61,0) (62,0) (63,0) (64,0)
    (65,0) (66,1) (67,0) (68,0) (69,0) (70,0) (71,0) (72,0)
    (73,0) (74,0) (75,1) (76,0) (77,0) (78,0) (79,0) (80,0)
    (81,0) (82,0) (83,0) (84,0) (85,0) (86,0) (87,0) (88,0)
    (89,0) (90,0) (91,0) (92,0) (93,0) (94,0) (95,0) (96,0)
    (97,0) (98,0) (99,0) (100,0) (101,0) (102,0) (103,0) (104,0)
    (105,0) (106,0) (107,0) (108,0) (109,0) (110,0) (111,0) (112,0)
    (113,0) (114,0) (115,0) (116,0) (117,0) (118,0) (119,0) (120,0)
    (121,0) (122,0) (123,0) (124,0) (125,0) (126,0) (127,0) (128,0)
    (129,0) (130,0) (131,0) (132,0) (133,0) (134,0) (135,0) (136,0)
    (137,0) (138,0) (139,0) (140,0) (141,0) (142,0) (143,0)
  };
  \addplot[fill=clrSearch] coordinates {
    (1,1) (2,2) (3,2) (4,5) (5,5) (6,5) (7,9) (8,9)
    (9,6) (10,7) (11,3) (12,5) (13,9) (14,12) (15,11) (16,5)
    (17,6) (18,6) (19,8) (20,5) (21,5) (22,12) (23,7) (24,7)
    (25,9) (26,8) (27,8) (28,7) (29,11) (30,18) (31,6) (32,12)
    (33,2) (34,8) (35,2) (36,4) (37,10) (38,9) (39,7) (40,7)
    (41,9) (42,9) (43,6) (44,10) (45,4) (46,8) (47,7) (48,11)
    (49,8) (50,8) (51,9) (52,5) (53,7) (54,8) (55,7) (56,5)
    (57,6) (58,6) (59,9) (60,7) (61,5) (62,6) (63,4) (64,2)
    (65,7) (66,5) (67,3) (68,10) (69,5) (70,6) (71,2) (72,3)
    (73,6) (74,3) (75,2) (76,5) (77,2) (78,1) (79,1) (80,6)
    (81,3) (82,1) (83,2) (84,3) (85,2) (86,0) (87,3) (88,1)
    (89,1) (90,4) (91,4) (92,2) (93,1) (94,2) (95,1) (96,1)
    (97,3) (98,2) (99,1) (100,0) (101,1) (102,0) (103,1) (104,0)
    (105,0) (106,0) (107,1) (108,1) (109,0) (110,1) (111,1) (112,0)
    (113,0) (114,0) (115,0) (116,0) (117,0) (118,0) (119,0) (120,0)
    (121,0) (122,0) (123,0) (124,0) (125,0) (126,0) (127,0) (128,0)
    (129,0) (130,0) (131,0) (132,0) (133,0) (134,0) (135,0) (136,0)
    (137,0) (138,0) (139,0) (140,0) (141,0) (142,0) (143,0)
  };
  \addplot[fill=clrRead] coordinates {
    (1,3) (2,5) (3,6) (4,12) (5,9) (6,6) (7,9) (8,11)
    (9,8) (10,6) (11,5) (12,5) (13,11) (14,13) (15,8) (16,11)
    (17,11) (18,10) (19,8) (20,11) (21,12) (22,13) (23,12) (24,10)
    (25,9) (26,10) (27,10) (28,8) (29,13) (30,11) (31,16) (32,13)
    (33,15) (34,11) (35,13) (36,10) (37,11) (38,11) (39,7) (40,9)
    (41,9) (42,7) (43,14) (44,12) (45,6) (46,18) (47,10) (48,10)
    (49,11) (50,12) (51,7) (52,9) (53,9) (54,10) (55,13) (56,13)
    (57,10) (58,12) (59,11) (60,10) (61,13) (62,10) (63,7) (64,7)
    (65,5) (66,8) (67,10) (68,5) (69,8) (70,8) (71,6) (72,7)
    (73,6) (74,7) (75,7) (76,4) (77,5) (78,10) (79,7) (80,3)
    (81,5) (82,9) (83,5) (84,5) (85,5) (86,9) (87,5) (88,6)
    (89,3) (90,3) (91,3) (92,2) (93,4) (94,2) (95,2) (96,3)
    (97,3) (98,2) (99,3) (100,3) (101,3) (102,1) (103,2) (104,2)
    (105,2) (106,2) (107,1) (108,0) (109,2) (110,1) (111,1) (112,1)
    (113,1) (114,1) (115,1) (116,1) (117,1) (118,1) (119,1) (120,0)
    (121,1) (122,1) (123,1) (124,1) (125,1) (126,1) (127,1) (128,1)
    (129,1) (130,1) (131,1) (132,1) (133,1) (134,1) (135,1) (136,1)
    (137,1) (138,1) (139,1) (140,1) (141,0) (142,0) (143,0)
  };
  \addplot[fill=clrOther] coordinates {
    (1,0) (2,2) (3,10) (4,7) (5,9) (6,2) (7,8) (8,6)
    (9,6) (10,9) (11,10) (12,6) (13,11) (14,8) (15,8) (16,4)
    (17,6) (18,5) (19,4) (20,8) (21,3) (22,6) (23,4) (24,5)
    (25,7) (26,11) (27,6) (28,4) (29,7) (30,9) (31,13) (32,7)
    (33,3) (34,9) (35,10) (36,8) (37,8) (38,5) (39,7) (40,4)
    (41,3) (42,7) (43,10) (44,7) (45,7) (46,6) (47,6) (48,7)
    (49,9) (50,7) (51,3) (52,5) (53,6) (54,3) (55,4) (56,5)
    (57,2) (58,4) (59,3) (60,5) (61,4) (62,5) (63,5) (64,4)
    (65,7) (66,9) (67,7) (68,7) (69,5) (70,6) (71,4) (72,5)
    (73,5) (74,4) (75,6) (76,4) (77,5) (78,4) (79,4) (80,5)
    (81,5) (82,6) (83,4) (84,4) (85,4) (86,7) (87,4) (88,4)
    (89,5) (90,4) (91,4) (92,3) (93,2) (94,1) (95,1) (96,1)
    (97,1) (98,1) (99,1) (100,1) (101,1) (102,0) (103,1) (104,1)
    (105,1) (106,1) (107,1) (108,0) (109,1) (110,1) (111,1) (112,0)
    (113,0) (114,0) (115,0) (116,0) (117,0) (118,0) (119,0) (120,1)
    (121,0) (122,0) (123,0) (124,0) (125,0) (126,0) (127,0) (128,0)
    (129,0) (130,0) (131,0) (132,0) (133,0) (134,0) (135,0) (136,0)
    (137,0) (138,0) (139,0) (140,0) (141,0) (142,1) (143,0)
  };
  \addplot[fill=clrFailed] coordinates {
    (1,26) (2,23) (3,19) (4,16) (5,14) (6,18) (7,18) (8,18)
    (9,18) (10,15) (11,19) (12,20) (13,14) (14,12) (15,17) (16,16)
    (17,16) (18,18) (19,16) (20,14) (21,15) (22,12) (23,14) (24,13)
    (25,14) (26,9) (27,12) (28,16) (29,11) (30,7) (31,8) (32,10)
    (33,11) (34,10) (35,7) (36,10) (37,8) (38,7) (39,9) (40,10)
    (41,10) (42,8) (43,7) (44,5) (45,10) (46,2) (47,7) (48,7)
    (49,5) (50,3) (51,8) (52,6) (53,3) (54,4) (55,3) (56,2)
    (57,6) (58,3) (59,4) (60,4) (61,3) (62,4) (63,4) (64,5)
    (65,3) (66,1) (67,1) (68,3) (69,3) (70,1) (71,4) (72,2)
    (73,1) (74,2) (75,0) (76,1) (77,3) (78,0) (79,1) (80,0)
    (81,0) (82,0) (83,1) (84,0) (85,0) (86,0) (87,0) (88,0)
    (89,1) (90,0) (91,0) (92,0) (93,0) (94,0) (95,0) (96,0)
    (97,0) (98,0) (99,0) (100,0) (101,0) (102,1) (103,0) (104,0)
    (105,0) (106,0) (107,0) (108,1) (109,0) (110,0) (111,0) (112,1)
    (113,0) (114,0) (115,0) (116,0) (117,0) (118,0) (119,0) (120,0)
    (121,0) (122,0) (123,0) (124,0) (125,0) (126,0) (127,0) (128,0)
    (129,0) (130,0) (131,0) (132,0) (133,0) (134,0) (135,0) (136,0)
    (137,0) (138,0) (139,0) (140,0) (141,1) (142,0) (143,1)
  };

  \end{groupplot}

  \node[
    anchor=north,
    yshift=-0.8cm,
    draw=black!40,
    fill=white,
    rounded corners=2pt,
    inner sep=3pt,
    font=\footnotesize,
  ] at ($(group c1r1.south west)!0.5!(group c3r1.south east)$) {%
    \begin{tikzpicture}[baseline=0.08cm]
      \def\sw{0.25cm}
      \def\sh{0.14cm}
      \def\gap{1.5cm}
      \fill[clrExplore] (0,0) rectangle (\sw,\sh);
      \node[anchor=west] at (\sw+0.05cm,0.07cm) {Explore};
      \fill[clrSearch] (\gap,0) rectangle (\gap+\sw,\sh);
      \node[anchor=west] at (\gap+\sw+0.05cm,0.07cm) {Search};
      \fill[clrRead] (2*\gap,0) rectangle (2*\gap+\sw,\sh);
      \node[anchor=west] at (2*\gap+\sw+0.05cm,0.07cm) {Read};
      \fill[clrOther] (3*\gap,0) rectangle (3*\gap+\sw,\sh);
      \node[anchor=west] at (3*\gap+\sw+0.05cm,0.07cm) {Other};
      \fill[clrFailed] (4*\gap,0) rectangle (4*\gap+\sw,\sh);
      \node[anchor=west] at (4*\gap+\sw+0.05cm,0.07cm) {Failed};
    \end{tikzpicture}%
  };

\end{tikzpicture}
}
\caption{Turn-wise action composition of DeepSeek-V4-Flash on Hard under T1 across the three agentic scaffolds.}
\label{fig:run_turns}
\end{figure*}
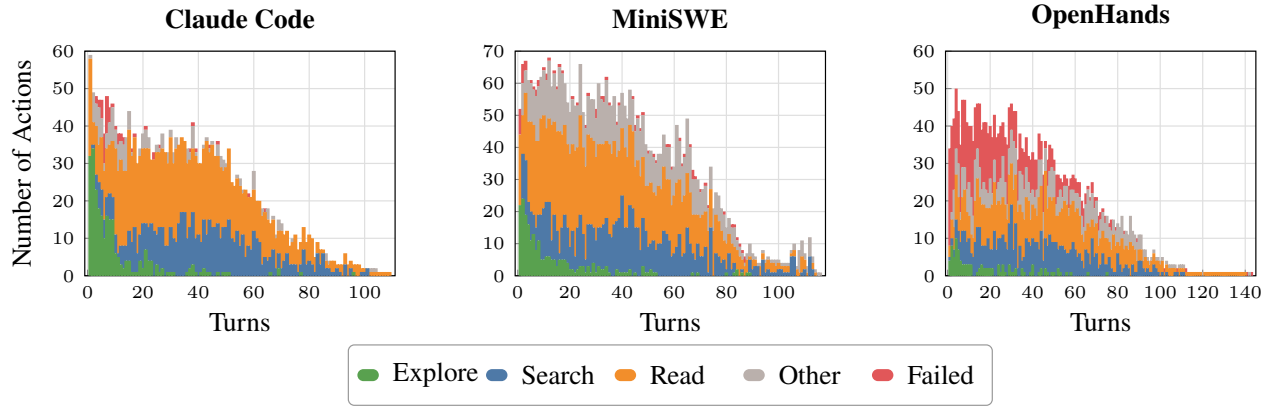

\pgfdeclarelayer{axis background}
\pgfsetlayers{axis background,main}

\definecolor{clrDSFlash}{HTML}{2166AC}
\definecolor{clrGLM}{HTML}{222222}
\definecolor{clrMiniMax}{HTML}{D6604D}
\definecolor{clrTierLow}{HTML}{CFE8F7}
\definecolor{clrTierMid}{HTML}{DCEEF8}
\definecolor{clrTierHigh}{HTML}{ECF5FA}
\definecolor{clrPareto}{HTML}{FE3621}
\definecolor{clrParetoLine}{HTML}{FE3621}
\definecolor{clrLegendGray}{HTML}{999999}

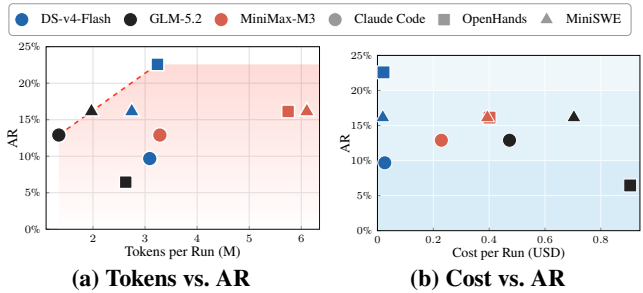
\begin{figure}[t]
\centering
\resizebox{\columnwidth}{!}{%
\begin{tikzpicture}
    \begin{axis}[
      hide axis,
      xmin=0, xmax=1, ymin=0, ymax=1,
      scale only axis,
      width=1pt, height=1pt,
      legend to name=scatterlegend,
      legend columns=6,
      legend style={
        font=\scriptsize,
        fill=white,
        draw=black!40,
        rounded corners=2pt,
        inner sep=3pt,
        column sep=1.5mm,
        row sep=1pt,
      },
    ]
      \addlegendimage{
        only marks, mark=*, mark size=3.6pt,
        draw=white, fill=clrDSFlash
      }
      \addlegendentry{DS-v4-Flash}
      \addlegendimage{
        only marks, mark=*, mark size=3.6pt,
        draw=white, fill=clrGLM
      }
      \addlegendentry{GLM-5.2}
      \addlegendimage{
        only marks, mark=*, mark size=3.6pt,
        draw=white, fill=clrMiniMax
      }
      \addlegendentry{MiniMax-M3}
      \addlegendimage{
        only marks, mark=*, mark size=3.6pt,
        draw=white, fill=clrLegendGray
      }
      \addlegendentry{Claude Code}
      \addlegendimage{
        only marks, mark=square*, mark size=3.2pt,
        draw=white, fill=clrLegendGray
      }
      \addlegendentry{OpenHands}
      \addlegendimage{
        only marks, mark=triangle*, mark size=4.2pt,
        draw=white, fill=clrLegendGray
      }
      \addlegendentry{MiniSWE}
    \end{axis}
    \node {\pgfplotslegendfromname{scatterlegend}};
\end{tikzpicture}%
}
\\[-1mm]
\begin{tabular}{@{}c@{\hspace{2mm}}c@{}}
  \resizebox{0.5\columnwidth}{!}{%
  \begin{tikzpicture}
    \pgfplotsset{
      oh point/.style={only marks, mark=square*,   mark size=4pt},
      cc point/.style={only marks, mark=*,         mark size=4.6pt},
      ms point/.style={only marks, mark=triangle*, mark size=5.6pt},
    }
    \begin{axis}[
      width=7.6cm, height=5.6cm,
      tick align=inside,
      grid=major,
      major grid style={line width=0.4pt, draw=gray!25},
      tick label style={font=\scriptsize},
      label style={font=\small},
      xlabel style={yshift=4pt},
      ylabel style={yshift=-6pt},
      title style={font=\small\bfseries},
      major tick length=2pt,
      xtick pos=bottom,
      ytick pos=left,
      axis line style={line width=0.6pt},
      every axis plot/.append style={draw=white, line width=0.9pt},
      xlabel={Tokens per Run (M)},
      ylabel={AR},
      xmin=1.1, xmax=6.35,
      ymin=0, ymax=25,
      xtick={2,3,4,5,6},
      ytick={0,5,10,15,20,25},
      yticklabel={\pgfmathprintnumber{\tick}\%},
    ]
      \path[
        shade,
        left color=clrPareto,
        right color=white,
        shading angle=0,
        opacity=0.18,
        draw=none,
      ]
        (axis cs:1.34395,0)
        -- (axis cs:1.34395,12.90)
        -- (axis cs:1.97089,16.13)
        -- (axis cs:3.23870,22.58)
        -- (axis cs:6.35,22.58)
        -- (axis cs:6.35,0)
        -- cycle;

      \addplot[
        dashed, mark=none, draw=clrParetoLine, line width=1pt, forget plot,
      ] coordinates {(1.34395,12.90) (1.97089,16.13) (3.23870,22.58)};

      \addplot[cc point, fill=clrDSFlash] coordinates {(3.08978,9.68)};
      \addplot[cc point, fill=clrGLM]     coordinates {(1.34395,12.90)};
      \addplot[cc point, fill=clrMiniMax] coordinates {(3.28443,12.90)};

      \addplot[oh point, fill=clrDSFlash] coordinates {(3.23870,22.58)};
      \addplot[oh point, fill=clrGLM]     coordinates {(2.62941,6.45)};
      \addplot[oh point, fill=clrMiniMax] coordinates {(5.74737,16.13)};

      \addplot[ms point, fill=clrDSFlash] coordinates {(2.74449,16.13)};
      \addplot[ms point, fill=clrGLM]     coordinates {(1.97089,16.13)};
      \addplot[ms point, fill=clrMiniMax] coordinates {(6.10663,16.13)};
    \end{axis}
  \end{tikzpicture}%
  }
  &
  \resizebox{0.48\columnwidth}{!}{%
  \begin{tikzpicture}
    \pgfplotsset{
      cc point/.style={only marks, mark=*,         mark size=4.6pt},
      oh point/.style={only marks, mark=square*,   mark size=4.2pt},
      ms point/.style={only marks, mark=triangle*, mark size=5.4pt},
    }
    \begin{axis}[
      width=7.6cm, height=5.6cm,
      tick align=inside,
      grid=major,
      major grid style={line width=0.5pt, draw=white, draw opacity=0.8},
      tick label style={font=\scriptsize},
      label style={font=\small},
      xlabel style={yshift=4pt},
      ylabel style={yshift=-6pt},
      title style={font=\small\bfseries},
      major tick length=2pt,
      xtick pos=bottom,
      ytick pos=left,
      axis line style={line width=0.6pt},
      every axis plot/.append style={draw=white, line width=0.6pt},
      xlabel={Cost per Run (USD)},
      ylabel={AR},
      xmin=0, xmax=0.94,
      ymin=0, ymax=25,
      xtick={0,0.2,0.4,0.6,0.8},
      ytick={0,5,10,15,20,25},
      yticklabel={\pgfmathprintnumber{\tick}\%},
    ]
      \begin{pgfonlayer}{axis background}
        \path[fill=clrTierLow,  opacity=0.8]
          (axis cs:0,0) rectangle (axis cs:0.94,10);
        \path[fill=clrTierMid,  opacity=0.8]
          (axis cs:0,10) rectangle (axis cs:0.94,20);
        \path[fill=clrTierHigh, opacity=0.8]
          (axis cs:0,20) rectangle (axis cs:0.94,25);
      \end{pgfonlayer}

      \addplot[cc point, fill=clrDSFlash] coordinates {(0.02634,9.68)};
      \addplot[cc point, fill=clrGLM]     coordinates {(0.47294,12.90)};
      \addplot[cc point, fill=clrMiniMax] coordinates {(0.22891,12.90)};

      \addplot[oh point, fill=clrDSFlash] coordinates {(0.02170,22.58)};
      \addplot[oh point, fill=clrGLM]     coordinates {(0.90513,6.45)};
      \addplot[oh point, fill=clrMiniMax] coordinates {(0.40146,16.13)};

      \addplot[ms point, fill=clrDSFlash] coordinates {(0.01930,16.13)};
      \addplot[ms point, fill=clrGLM]     coordinates {(0.70347,16.13)};
      \addplot[ms point, fill=clrMiniMax] coordinates {(0.39361,16.13)};
    \end{axis}
  \end{tikzpicture}%
  }
  \\[-1mm]
  \phantomsubcaption\label{fig:cost_tokens}%
  {\small\bfseries (a) Tokens vs.\ AR}
  &
  \phantomsubcaption\label{fig:cost_usd}%
  {\small\bfseries (b) Cost vs.\ AR}
\end{tabular}
\caption{Efficiency--effectiveness trade-offs among frontier coding agents on Hard under T1. The two panels compare AR with tokens and cost per run, respectively.}
\label{fig:cost}
\end{figure}

Beyond detection effectiveness, we further evaluate the practical efficiency of coding agents.
We examine how per-run resource consumption relates to end-to-end vulnerability detection.
As shown in Figure~\ref{fig:cost}(\subref{fig:cost_tokens}), the token--AR Pareto frontier extends from Claude Code with GLM-5.2 to OpenHands with DeepSeek-V4-Flash.
Since API pricing varies across models, Figure~\ref{fig:cost}(\subref{fig:cost_usd}) further compares per-run cost with AR.
OpenHands with DeepSeek-V4-Flash achieves the highest AR while remaining among the least expensive configurations, whereas OpenHands with GLM-5.2 incurs the highest cost but obtains the lowest AR.
Overall, higher resource consumption does not necessarily yield more effective vulnerability discovery.

\subsection{Agent Trajectory Analysis}

To characterize agent behavior across interaction turns, we analyze action distributions for DeepSeek-V4-Flash.
For this analysis, we categorize the provided commands into three types of actions: \textbf{\textcircled{1} Explore} commands (\texttt{find}, \texttt{ls}), \textbf{\textcircled{2} Search} commands (\texttt{rg} and \texttt{grep}), and \textbf{\textcircled{3} Read} commands (\texttt{cat}, \texttt{head}, \texttt{sed}, \texttt{tail}).
\textbf{\textcircled{4} Other} denotes actions beyond those explicitly provided to the agent in the prompt.

As shown in Figure~\ref{fig:run_turns}, Read accounts for the largest share of actions across all configurations, while Explore is concentrated in the initial turns and Search persists throughout the trajectories.
The three scaffolds also exhibit distinct action patterns.
Claude Code produces comparatively few Failed actions and also executes few Other actions.
MiniSWE exhibits a larger share of Other actions, corresponding to commands beyond those specified in the prompt.
By contrast, OpenHands produces substantially more Failed actions, particularly during the early turns.

\section{Conclusion} \label{sec:conclusion}
    We presented VulnGym, a repository-level benchmark for evaluating vulnerability detection by coding agents.
It contains 184 advisories and 408 vulnerability entries
, with fine-grained annotations of entry points, critical operations, and vulnerability traces.
Based on these annotations, VulnGym combines an end-to-end task with three oracle-based subtasks to assess overall performance and diagnose capability limitations.
Our evaluation shows that
challenging vulnerabilities remain largely unresolved and performance varies across LLM--agent combinations.
With this benchmark,
VulnGym provides a foundation for advancing coding agents toward reliable repository-level vulnerability detection.

\end{sloppypar}

\bibliography{aaai2027}

\end{document}